\documentclass[aps, twocolumn, prb, showpacs, amsmath, amssymb, floatfix]{revtex4-1}
\newcommand{\ket}[1]{\left|#1\right>}

\usepackage{graphicx}
\usepackage{dcolumn}
\usepackage{bm}
\usepackage{amsmath}
\usepackage{times}
\usepackage[usenames]{color}
\usepackage{natbib}
\usepackage{ulem}

\begin{document}
\title{Photoinduced collective mode, inhomogeneity, and melting in a charge-order system}
\author{Hitoshi Seo$^{1,2}$, Yasuhiro Tanaka$^{3}$, and Sumio Ishihara$^4$}
\affiliation{$^1$Condensed Matter Theory Laboratory, RIKEN, Wako 351-0198, Japan}
\affiliation{$^2$Center for Emergent Matter Science (CEMS), RIKEN, Wako 351-0198, Japan}
\affiliation{$^3$Department of Physics, Chuo University, Tokyo 112-8551, Japan}
\affiliation{$^4$Department of Physics, Tohoku University, Sendai 980-8578, Japan}
\date{\today}
\begin{abstract} 
We theoretically investigate photoresponses of a correlated electron system upon stimuli of a pulsed laser light. 
Real-time dynamics of an interacting spinless fermion model on a one-dimensional chain, 
as a model of charge order (CO),  
are numerically simulated using the time-dependent Hartree-Fock method. 
In particular, we discuss the differences between two situations as the initial state: 
 the homogeneous order 
 and the presence of a domain wall, i.e., a kink structure embedded in the CO bulk. 
Coherent dynamics are seen in the former case: 
When the frequency of the pump light $\omega_{p}$ is varied, 
along with single particle excitations across the CO gap ($\Delta_\textrm{CO}$),  
the resonantly-excited collective phase mode
 near $\omega_{p} \simeq \Delta_\textrm{CO}/2$ efficiently destabilizes CO. 
In clear contrast, in the latter case, when $\omega_{p}$ is tuned at such in-gap frequencies 
 and the intensity of light is sufficiently large, 
 inhomogeneity spreads out from the kink to the bulk region through kink creations. 
Moreover, even stronger intensity induces the inhomogeneous melting of CO where the CO gap is destroyed. 
\end{abstract} 
%
%
%
%

\maketitle

\section{Introduction}
Photoresponses of condensed matter are now increasingly upon intense research 
 thanks to rapid advancements in technologies~\cite{Orenstein,Kampfrath}. 
Ultrafast spectroscopies have been applied to 
a wide variety of correlated systems, 
 whose displayed physical phenomena include 
 Mott metal-insulator transition~\cite{Iwai}, 
 charge order (CO)~\cite{Fiebig}, 
 charge-density-wave~\cite{Chollet,Tomaljak}, 
 superconductivity~\cite{Matsunaga},  
 and magnetism~\cite{Ogasawara}. 
The understanding of their dynamical responses upon pulsed laser light emission 
 is of fundamental interest owing to its nonequilibrium and nonstationary nature, 
 which is mostly unexplored for correlated matters. 
Elucidating their mechanisms is crucial also for possible future applications using phase control. 
In fact, in certain materials
 the light irradiation induces macroscopic conversion of phases; 
 the phenomenon called photoinduced phase transition (PIPT)~\cite{Nasu,Tokura} occurs.
Among them, molecular crystals have been intensively studied, 
 taking advantage of their lower energy scale 
 which in turn results in slower time scale  
 compared to other materials, for example, transition metal oxides. 
Such a feature enables us to access the early stages leading to PIPT, 
 and even the initial processes governed by electronic degrees of freedom have started to become visible.\\
\indent 
Experimentally, 
 the existence of spatial inhomogeneity is often indicated to 
 emerge through the PIPT process, 
 and is discussed to be an essential ingredient for the realization of PIPT, 
 referred to as the domino effect~\cite{Koshihara}. 
A well known example is the case of quasi-one-dimensional mixed stack compound TTF-CA [tetrathiafulvalene-p-chloranil]
 which shows a valence transition by the collective charge transfer between two kinds of molecules TTF and CA. 
The light irradiation to the low temperature ionic phase leads to a transient neutral phase, 
 whose time-resolved measurements revealed domain wall excitations and growth dynamics~\cite{Iwai_TTFCA}. 
Such inhomogeneous dynamics have been discussed also in quasi-two-dimensional 
 CO materials based on the BEDT-TTF [bis(ethylenedithio)tetrathiafulvalene] molecule. 
Depending on the nature of their equilibrium CO transition, 
 the dynamic domain formation and how they evolve turn out to be material dependent
 and reflect the ability of CO melting~\cite{Iwai_BEDTTTF}.\\ 
\indent 
Theoretical efforts to elucidate the early stage dynamics of PIPT phenomena 
 have been made using microscopic models of correlated electrons, 
 often including electron-phonon couplings which eventually lead to structural changes at longer time scales. 
In such calculations, 
 coherent (spatially homogeneous) dynamics have been mainly studied, while
 theoretical treatments of interactions fully taking into account quantum fluctuation 
 limit the system size we can treat~\cite{Yonemitsu,Takahashi,Dagotto,Hashimoto,Hashimoto2}. 
Then, to discuss the role of inhomogeneity mentioned above, 
 different approaches have been used, e.g.,
 implementing by hand the existence of domain structures, randomness, or local excitation~\cite{Nagaosa,Miyashita,Iwano,Tanaka}. 
The inter-relationship between the two different ways for phase conversion, 
 i.e., coherent excitation and inhomogeneity,  
 in the initial processes remains to be clarified, 
 which is indispensable for the microscopic understanding of PIPT.\\
\indent In this paper, 
 we propose a unified picture treating the above-mentioned two processes on equal footing
 by means of a microscopic model of interacting particles. 
For this purpose, 
 we choose a minimal model of a correlated electron system for CO 
 and apply the time-dependent Hartree-Fock (HF) method, 
 whose combination enables us to treat system sizes large enough 
 for a systematic survey including inhomogeneity in the real-time dynamics. 
We find that the coherent dynamics can be largely enhanced by tuning the pump-light frequency 
 at the in-gap region of the CO gap.  
This is interpreted as collective modes associated with the CO, bringing about large oscillations of the order parameter. 
When a kink structure, i.e., a domain wall of CO phases, is present in the initial state, 
 we show that the inhomogeneity emerges starting from this kink 
 when the collective excitations are activated and eventually results in CO melting 
 at large enough intensity of the light.\\ 
\section{Model and method}
We treat an interacting spinless fermion model on a one-dimensional chain, 
 whose time ($\tau$) dependent Hamiltonian is written as 

\begin{align}
\mathcal{H}(\tau)=
&\sum_{i}  \left[ t_\textrm{nn} (\tau)\ c_{i}^{\dagger}c_{i+1}+\text{h.c.} \right] \notag \\
&+ \sum_{i} \left[ t_\textrm{nnn} (\tau)\ c_{i}^{\dagger}c_{i+2}+\text{h.c.} \right] 
+V\sum_{i} n_{i} n_{i+1}, 
\label{eq:hamil}
\end{align}

\noindent where $c_{i} (c_{i}^{\dagger})$ is the annihilation (creation) operator of a spinless fermion at site $i$,
and $n_{i} \equiv c_{i}^{\dagger}c_{i}$ is the fermion number operator. 
The first two terms are the kinetic energy with hopping integrals between 
 site pairs of nearest neighbor and next nearest neighbor, 
 and the last term is the nearest neighbor Coulomb repulsion which induces CO at half filling.  
The effect of external laser pulse polarized along the chain direction 
 is incorporated by the Peierls substitution in the hopping integrals as 
\begin{align}
t_\textrm{nn}(\tau)= t\ e^{iA(\tau)}, \hspace{2mm} t_\textrm{nnn}(\tau)=t' e^{2iA(\tau)},
\end{align}
where the lattice constant, the light velocity, the elementary charge, and $\hbar$ 
are taken as unity.  
We consider the vector potential for the pump light centered at $\tau=0$ 
 with a gaussian envelope of frequency $\omega_{p}$ 
\begin{align}
A(\tau)=\frac{A_{p}}{\sqrt{2\pi}\tau_{p}}\exp{\left(-\frac{\tau^{2}}{2\tau_{p}^{2}}\right)}\cos(\omega_{p}\tau),
\label{eq:vectorpotential}
\end{align}
where $A_{p}$ and $\tau_{p}$ control the amplitude and the pulse width, respectively. 
In the following we set $t=1$ as the unit of energy (and time $1/t$), and fix $t'=0.1$ and $V=4$; 
 we also fix $\tau_{p}=5$. 
Then the parameters we vary are $\omega_{p}$ and $A_{p}$.   

For the time evolution, a 
 time-dependent HF method at absolute zero temperature 
 developed in Refs.~\cite{Terai,Kuwabara,Tanaka} is adopted. 
In this method the interaction term is treated within the HF approximation:
$n_i n_{i+1} \rightarrow \langle n_i \rangle n_{i+1} + n_i \langle n_{i+1} \rangle - \langle n_i \rangle \langle n_{i+1} \rangle
- \langle c_{i}^{\dagger}c_{i+1} \rangle c_{i+1}^{\dagger}c_{i} - c_{i}^{\dagger}c_{i+1} \langle c_{i+1}^{\dagger}c_{i} \rangle 
+ \langle c_{i}^{\dagger}c_{i+1} \rangle \langle c_{i+1}^{\dagger}c_{i} \rangle$. 
After evaluating the mean-fields for all sites and bonds in a self-consistent manner at equilibrium, 
 the time evolution of the wave vector $\ket{\Psi(\tau)}$ is calculated by the discretized Schr$\rm \ddot o$dinger equation as 
\begin{align} 
\ket{\Psi(\tau+\Delta\tau)} = 
 \exp{\left[-i \ {\cal H}^{HF}(\tau+\frac{\Delta\tau}{2}) \Delta\tau \right]} \ \ket{\Psi(\tau)}.
\end{align}
The transient Hamiltonian ${\cal H}^{HF}(\tau+\Delta\tau/2)$ is calculated by 
 a linear interpolation 
\begin{align} 
{\cal H}^{HF}(\tau+\frac{\Delta\tau}{2})= \frac{1}{2}\left[ {\cal H}^{HF}_0(\tau+\Delta\tau) +  {\cal H}^{HF}(\tau) \right], 
\end{align}
 where ${\cal H}^{HF}_0(\tau+\Delta\tau)$ is the ``extrapolated'' HF Hamiltonian 
 with mean fields calculated by the wave vector 
\begin{align} 
\ket{\Psi^{0}(\tau+\Delta\tau)} = 
 \exp{\left[-i \ {\cal H}^{HF}(\tau) \Delta\tau  \right]} \ \ket{\Psi(\tau)}.
\end{align}
This scheme gives the accuracy in $\ket{\Psi(\tau+\Delta\tau)}$ with errors of the order $(\Delta\tau)^3$, 
 and considerably lightens the numerical calculations. 
Note that by using HF approximation, 
 we are implying higher dimension (quasi-one-dimensional) systems in mind, 
 rather than trying to investigate a purely one-dimensional system, as studied in Ref.~\cite{Hashimoto3}. 

\section{results}\label{sec_results}
In the following, we show results for the system sizes $N=500$ at half filling (Sec. \ref{subsec_homo})
 and $N=501$ with one fermion added to that (Sec. \ref{subsec_inhomo}), 
 using the periodic boundary condition. 
These sizes are large enough to neglect finite-size effect within the time domain we consider, 
 typically $\tau \leq 100$. 
\subsection{Homogeneous order}\label{subsec_homo}
\begin{figure}[htb]
\begin{center}
\includegraphics[width=7cm]{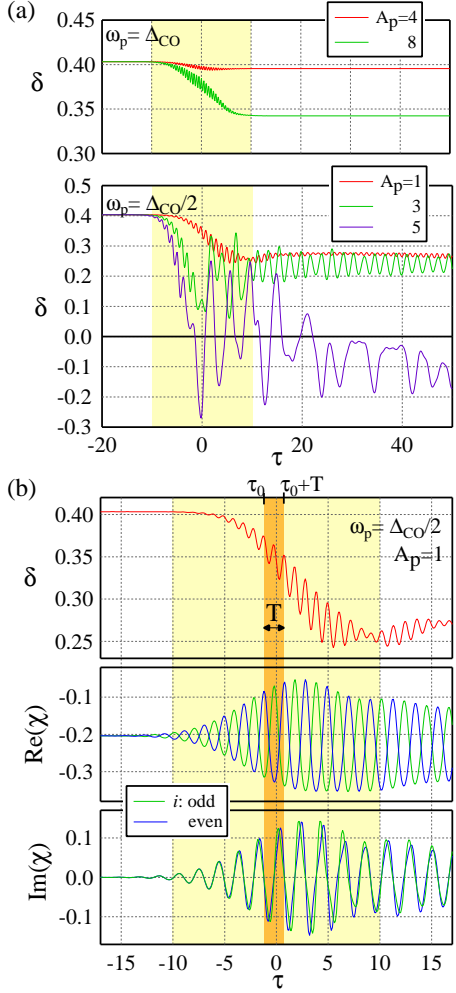}
\end{center}
\vspace{-1cm}
\caption{ 
(a) $\tau$ dependence of order parameter $\delta(\tau)$ for an even number of sites, with two pump-light frequencies: 
$\omega_p = \Delta_\textrm{CO} = 6.45$ and $\omega_p = \Delta_\textrm{CO}/2$, 
 for different intensity $A_p$. 
(b) A closer look at the latter case for $A_p=1$ together with real and imaginary parts of the bond orders $\chi_i(\tau)$
 for even and odd $i$. 
The light and dark shaded areas show the time domain where the pump light is applied and one period of oscillation, respectively.
}
\label{fig1}
\end{figure}

First, we show results for even number of sites ($N=500$). 
In this case the twofold periodicity $\langle n_i \rangle = 1/2 + (-1)^i \delta(\tau)$ 
 is always maintained throughout the simulation, namely, coherent dynamics are seen~\cite{note_randomness}. 
In the initial state $\tau=\tau_\textrm{in}$, our parameter set  
 gives the order parameter $\delta(\tau_\textrm{in})=0.40$ $(\equiv \delta_0)$ and 
 the one-particle gap $\Delta_\textrm{CO}=6.45$. 
In Fig.~\ref{fig1}(a), 
 we show the $\tau$ dependence of $\delta(\tau)$ for two pump-light frequencies: 
 $\omega_p = \Delta_\textrm{CO}$ and $\omega_p =\Delta_\textrm{CO}/2$. 
When $\omega_p \gtrsim \Delta_\textrm{CO}$, during the pump light is shined $\delta$ decreases 
 and stays constant afterwards. 
This is interpreted as the one-particle excitation where the energy of light is 
 absorbed by the fermion system through photocarrier generation~\cite{Hashimoto3,Ishihara}. 

On the other hand, the response is larger for $\omega_p \simeq \Delta_\textrm{CO}/2$, 
 where the decrease in $\delta(\tau)$ is large and coherent oscillations are seen even after the light irradiation is finished. 
This is depicted in more detail in Fig.~\ref{fig1}(b), 
 where $\delta(\tau)$ together with the real and imaginary parts of the bond orders 
 $\chi_i(\tau)=\langle c_{i}^{\dagger}c_{i+1}\rangle$, for even and odd $i$, are plotted 
 for $\omega_p = \Delta_\textrm{CO}/2$, $A_p=1$. 
Since the gauge field induces electric current with frequency $\omega_p$, 
 the imaginary part of $\chi_i(\tau)$
follows its time profile, in the in-phase way for even and odd $i$. 
This directly generates an oscillation in the charge density in a coherent way. 

We can write the modulation in $\langle n_i \rangle$ as $- (-1)^i \Delta\delta(\tau)$ 
 where the decrease in the order parameter $\Delta\delta(\tau)$ is approximately written as 
 $\Delta\delta(\tau) \propto \cos{\left[\alpha \cos{\omega_p \left(\tau-\tau_0\right)}\right]}$, 
 with $\alpha$ being a coefficient representing the oscillation in time.
Then the modulation is rewritten as 
 $- (-1)^i \Delta\delta(\tau) \propto - \cos{\left[\pi i + \alpha \cos{\omega_p \left(\tau-\tau_0\right)}\right]}$. 
This is the collective phase mode out of CO ground state\cite{Fukuyama}. 
As shown in Fig.~\ref{fig1} (b), within one period $T = 2\pi/\omega_p$, 
 the induced modulation in $\langle n_i \rangle$ undergoes two oscillations, namely,  
 it has the frequency $2\omega_p=\Delta_\textrm{CO}$. 

In such a picture, 
 the real part of $\chi_i(\tau)$ shows a modulation proportional to 
 $\cos{\left[\pi \left(i+1/2\right) + \beta\ \cos{\omega_p \left(\tau-\tau_0\right)}\right]}$, 
 where $\beta$ is a coefficient as above; 
 $i+1/2$ is the midpoint coordinate between neighboring sites. 
This is equal to $-\sin{\left[\pi i + \beta\ \cos{\omega_p \left(\tau-\tau_0\right)}\right]}$ with period $T$ in time, 
 consistent with the numerical results in Fig.~\ref{fig1}(b). 
We note that such in-gap modes are observed in other systems 
 such as 1/2-filled models~\cite{Lu} and models for BCS superconductors~\cite{Matsunaga,Krull}
whose origins are discussed to be two-photon processes. 
In clear contrast, in our case the excitation is an optically-allowed process 
 seen in the linear absorption spectra in the small $A_p$ limit (see Appendix \ref{AppA}). 

\begin{figure}[htb]
\begin{center}
\includegraphics[width=9.2cm]{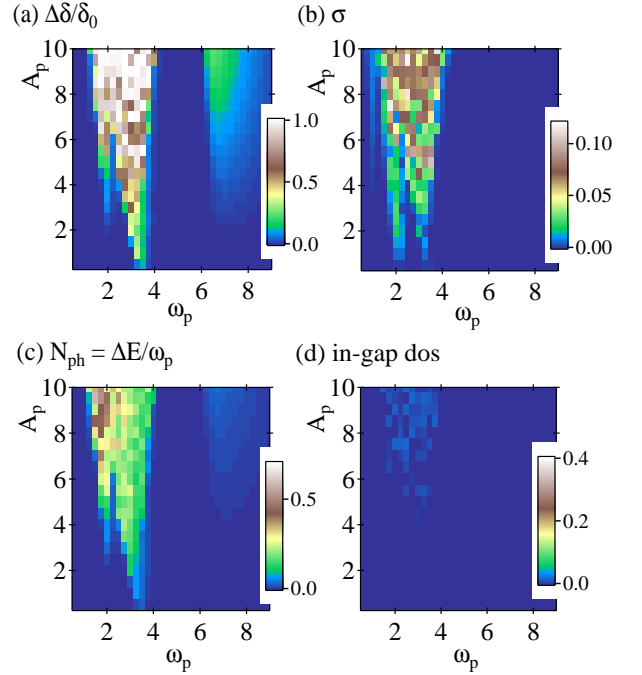}
\end{center}
\vspace{-0.7cm}
\caption{(Color online) 
Two-dimensional plots on the \{$\omega_p, A_p$\} plane of the time averaged photoresponses for the $N=500$ sites case: 
(a) decrease in the order parameter $\Delta\delta$ 
 renormalized by $\delta_0$, 
(b) standard deviation $\sigma$ in $\delta(\tau)$, 
 (c) effective absorbed photon number $N_{ph}$, 
 and (d) in-gap density of states (dos). 
Their definitions are depicted in the text.
}
\label{fig2}
\end{figure}

We can see the resonant character 
in the two-dimensional plots on the \{$\omega_p, A_p$\} plane. 
In Fig.~\ref{fig2}, we plot (a) the time averaged decrease in the order parameter 
 $\Delta\delta= \langle \Delta\delta(\tau) \rangle_\textrm{av}$ renormalized by $\delta_0$, 
 (b) the standard deviation 
 $\sigma = \langle \sqrt{(\delta(\tau) - \bar{\delta})^2} \rangle_\textrm{av}$ 
 where $\bar{\delta} =  \langle \delta(\tau) \rangle_\textrm{av}$, 
 a measure for the oscillatory part, 
 (c) the effective absorbed photon number $N_{ph}$ defined as $\Delta E / \omega_p$ where $\Delta E$ 
 is the increase in the expectation value of energy, 
 and (d) the in-gap density of states (dos) measured within $\pm 0.1t$ range of the Fermi level, 
 compared to that in the $V=0$ equilibrium metallic state. 
$\Delta\delta/\delta_0$, $\sigma$, $\bar{\delta}$, and the in-gap density of states are the values averaged over $\tau=20$ to 60, 
 and $\Delta E$ is converged after the light emission in our simulation.
We note that negative values of $\delta(\tau)$ are observed in some cases for large $A_p$ 
 as seen in Fig.~\ref{fig1}(a). This indicates that the phase of CO is reverted. 
For the cases where the time average $\bar{\delta}$ is negative, 
 we take its absolute value $| \bar{\delta} |$ to calculate the time averaged decrease, 
 as $\Delta\delta=\delta_0 - | \bar{\delta} |$. 

The excitation across the gap is 
seen for $\omega_p \gtrsim \Delta_\textrm{CO}$ in $\Delta\delta$ and $N_{ph}$ in Figs.~\ref{fig2}(a) and 2(c), respectively,
 while $\sigma$ is negligibly small there, as seen in Fig~\ref{fig2}(b). 
On the other hand, the large decrease $\Delta\delta$ occurs resonantly as in-gap excitations, 
 together with appreciable $\sigma$ indicating its oscillating nature. 
In the small $A_p$ region, the resonance frequency approaches $\omega_p \simeq \Delta_\textrm{CO}/2$, 
 whereas the mode broadens and enlarges as $A_p$ is increased. 
This is because the pulsed laser light includes Fourier components of  different frequencies 
 and also the nonlinear effects become prominent. 
We note the existence of another mode at $\omega_p \simeq 2$, 
 separated from the phase mode in the small $A_p$ region of Fig.~\ref{fig2}, 
 which is the amplitude mode excitation of CO. 
This is more clearly seen in $\sigma$ than in $\Delta\delta$, 
 indicating that it does not efficiently acts to decrease $\delta$, 
 compared to the phase mode (see Appendix \ref{AppB} for details). 

\begin{figure*}[tb]
\begin{center}
\includegraphics[width=18.5cm]{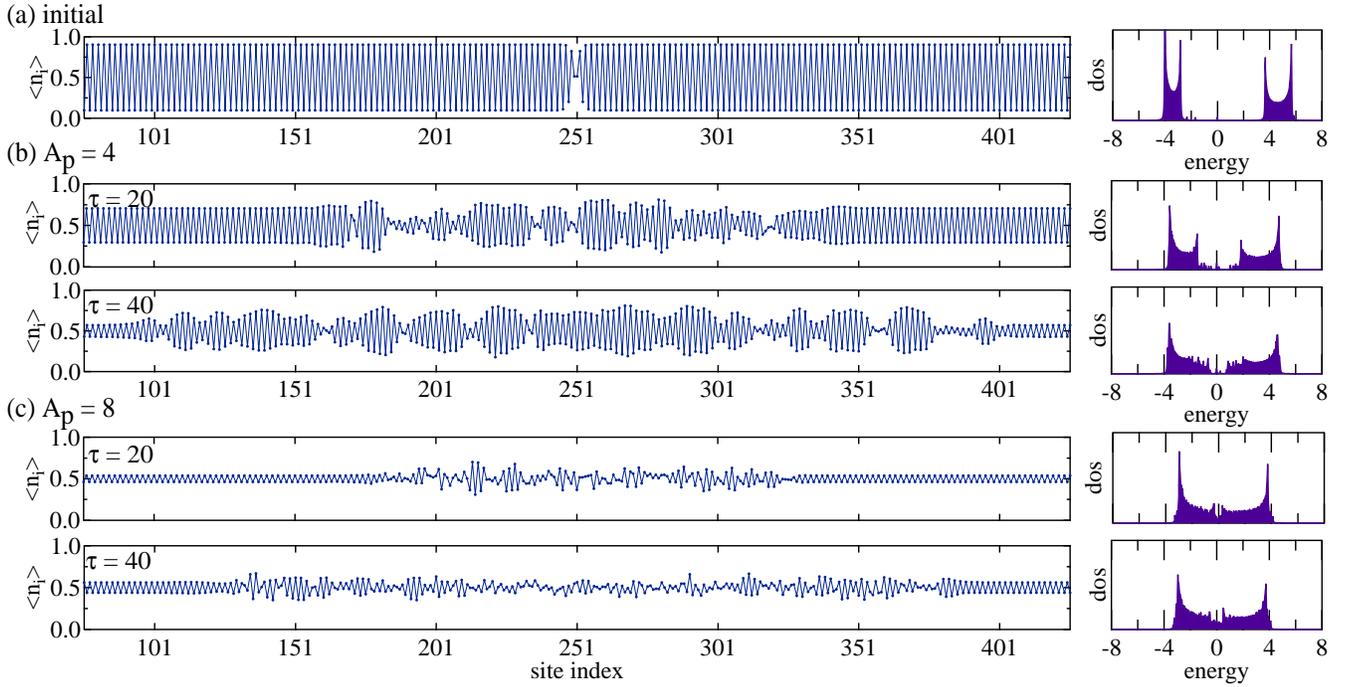}
\end{center}
\vspace{-0.7cm}
\caption{(Color online) 
Time evolution of $\langle n_i \rangle$ on each site, plotted for $i=$ 76 to 426, among $N=$ 501 sites.
(a) The initial state having a kink structure centered at $i=$ 250-251. 
(b) $A_p=$ 4 and (c) $A_p=$ 8, for time $\tau=$ 20 and 40 [$\omega_p = \Delta_\textrm{CO}/2$]. 
On the right of each plot, the temporary density of states (dos) is plotted, where the Fermi energy is set to the origin.  
}
\label{fig3}
\end{figure*}
These modes become mixed in the large $A_p$ regime 
 and therefore the photoresponses show complex $\tau$ dependence 
 with different frequency components difficult to resolve [see Fig.~\ref{fig1}(a)]. 
The in-gap density of states shown in Fig.~\ref{fig2}(d) is negligibly small in all parameter space, 
 despite the fact that resonant excitation sometimes accompanies phase inversion of CO 
 as mentioned above, 
 leading to instantaneous disappearance of the order [$\delta(\tau)=0$], 
 for example seen in Fig.~\ref{fig1} (a) for $A_p=5$ 
 (see Appendix \ref{AppBB} for examples of temporary density of states). 
This will be contrasted with the case of inhomogeneous dynamics in the next subsection.

\subsection{Presence of a domain wall}\label{subsec_inhomo}
\begin{figure}[tb]
\begin{center}
\includegraphics[width=8cm]{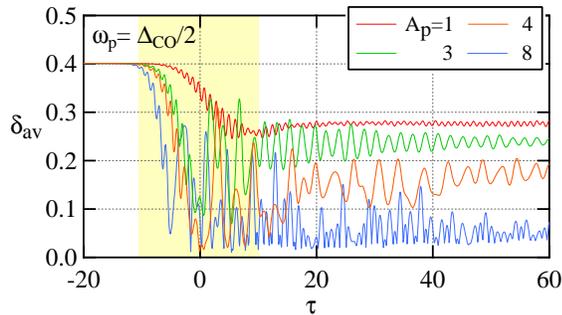}
\end{center}
\vspace{-0.5cm}
\caption{(Color online) 
$\tau$ dependence of the averaged ``order parameter'' $\delta_\textrm{av}(\tau)$ 
 for $N=501$; 
The frequency of the pump light is $\omega_p = \Delta_\textrm{CO}/2$, 
 for different intensity $A_p$. 
The shaded area show the time domain where the pump light is applied. 
}
\label{fig4}
\end{figure}

\begin{figure}[tb]
\vspace{-0.3cm}
\begin{center}
\includegraphics[width=9.2cm]{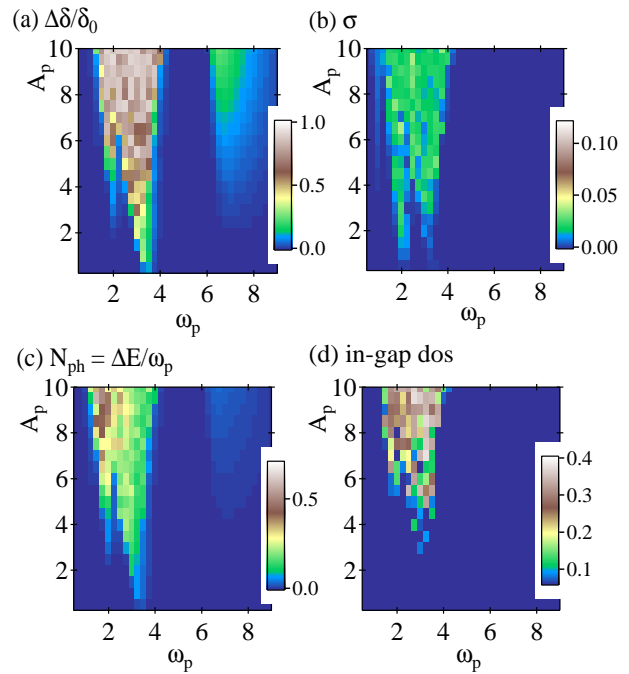}
\end{center}
\vspace{-0.7cm}
\caption{(Color online) 
Two-dimensional plots on the \{$\omega_p, A_p$\} plane of the time averaged photoresponses for the $N=501$ sites case: 
(a) decrease in the order parameter $\Delta\delta$ 
 renormalized by $\delta_0$, 
(b) standard deviation $\sigma$ in $\delta(\tau)$, 
(c) effective absorbed photon number $N_{ph}$, 
 and (d) in-gap density of states (dos).
Their definitions are depicted in the text.
}
\label{fig5}
\end{figure}

Next, we show results for odd number of sites ($N=501$). 
Since the twofold periodicity of CO does not fit to the system, 
 the equilibrium self-consistent HF solution shows a kink structure~\cite{OnoTerai} as a domain wall embedded in the CO bulk. 
This is shown in Fig.~\ref{fig3}(a), where the real-space plot of charge density on each site $\langle n_i \rangle$ is shown. 
For $\omega_p \gtrsim \Delta_\textrm{CO}$, 
 the response is unchanged from the $N$ even case: 
 The bulk region shows coherent dynamics irrespective of the presence of the kink.\\ 
\indent
When the pump light frequency is tuned in the in-gap region, 
 especially near $\omega_p \simeq \Delta_\textrm{CO}/2$ as in the $N$ even case,
 a large response is seen again. 
When $A_p$ is small the response is similar to the $N$ even case, 
 namely, the collective phase mode is resonantly excited and the system remains coherent. 
However, 
 appreciable differences are observed in the large $A_p$ regime: Two distinct types of inhomogeneous dynamics are seen, 
 as 
 represented in the time evolution of $\langle n_i \rangle$ 
 plotted for $\tau=20$ and 40 (``snapshots''), 
 in Figs.~\ref{fig3}(b) and 3(c) for $A_p=4$ and 8, respectively~\cite{Suppl}. 
At first sight, as a common feature, 
 one can see the generation of inhomogeneous regions in both cases, 
 and its growth as time evolves originated from the kink in the initial state. 
In the following we discuss qualitatively different features of the two cases.\\ 
\indent
In the former case of $A_p=4$ [Fig.~\ref{fig3}~(b)],  
 the inhomogeneity is accompanied by kink generations.  
This is reflected in the $\tau$ dependence of the averaged ``order parameter'' $\delta_\textrm{av}(\tau)$, 
 defined as $\delta_\textrm{av}(\tau) = (2N)^{-1} \sum_i | \langle n_i \rangle -  \langle n_{i+1} \rangle|$ 
 shown in Fig.~\ref{fig4}; $\delta_\textrm{av}(\tau)$ coincides with $\delta(\tau)$ in the homogeneous case. 
As $A_p$ is increased, first the curve is almost identical to the $N=$ 500 case 
 [see data for $A_p=1$ in Fig.~\ref{fig2}~(a)]. 
In the data for $A_p=$ 3 and 4, the large oscillations in $\delta_\textrm{av}(\tau)$ are seen in the initial stage 
 and then they decay gradually. 
These behaviors come from the phase mode excitation in the bulk CO region 
 and the growth of inhomogeneous region reducing such oscillating CO, respectively.
On the other hand, for the latter case with a larger value of $A_p=8$ [Fig.~\ref{fig3}(c)], 
 in the inhomogeneous region the CO structure is nearly lost; a CO melting occurs. 
In this case, $\tau$ dependence of $\delta_\textrm{av}(\tau)$ shows a very different curve 
 as seen in Fig.~\ref{fig4}; 
 the oscillatory behavior is almost diminished at some time after the laser radiation 
 and the value of $\delta_\textrm{av}(\tau)$ approaches to a very small value ($\simeq 0.04$ in this case).\\ 
\indent
The difference of the two cases is highlighted in the temporal evolution of the density of states, 
 plotted in the right panels in Fig.~\ref{fig3} (see also Ref. \cite{Suppl}). 
One can see that for $A_p=4$ the system essentially maintains the gap structure due to CO, 
 reflecting the fact that the CO structure in real space is basically kept even in the presence of multiple kinks 
 which give rise to in-gap states. 
For $A_p=8$, in contrast, the gap structure itself gets diminished. 
The shape of the density of states becomes similar to that for the metallic state in equilibrium at $V=0$ 
 but with heavy influence of disorder.
The transfer integrals are renormalized in the case of $V\neq0$, 
 owing to the Fock term, as 
 $t \rightarrow \tilde{t} \equiv t - V \chi_i(\tau)$. 
The calculated $\chi_i(\tau)$ typically take values such as $-0.1 \sim -0.4$, 
therefore $\tilde{t} \approx 1.2t \sim 2.6t$. 
Moreover, the Hartree term gives rise to modulations in the on-site potentials 
 of $V (\langle n_i \rangle -0.5) \approx \pm 0.8t$ at maximum. 
These two randomness effects result in the broad and blurred shapes of the density of states in Fig.~\ref{fig3}(c).\\ 
\indent
In Fig.~\ref{fig5}, we show the two-dimensional plots on the \{$\omega_p, A_p$\} plane, 
 for the same quantities as in Fig.~\ref{fig2}, also averaged over $\tau=20$ to 60. 
One can see the response is large in the same in-gap region as in the $N$ even case, 
 indicating that the resonant collective mode excitation is an essential ingredient. 
The difference from the $N$ even case is prominent in the in-gap density of states in Fig.~\ref{fig5}(d): 
There is a threshold behavior for the raise, 
 around $A_p=3$ - $4$ for the resonant condition $\omega_p = \Delta_\textrm{CO}/2 \simeq 3.2$, 
 where the kink generation and inhomogeneous growth start to occur. 
The transformation to the ``CO melting'' behavior is rather obscure and not well defined but 
 at $A_p=7$ - $8$ and larger, the in-gap density of states becomes large and a metallic like state is achieved. 
These are in clear contrast with the $N$ even case, where such in-gap density of states are always negligibly small [Fig.~\ref{fig2}(d)]. 

We note that, in both cases, the inhomogeneity is generated by 
 the cooperative effect of the phase mode excitation and the existence of a kink in the middle working as a seed. 
As $A_p$ increases, the former brings about the reduction of CO order parameter 
 together with the large oscillation in the bulk CO region, 
 sometimes even accompanying its sign change. 
This makes instantaneous disappearances of CO and enables the kink generations and melting possible. 
In fact, 
the growth of inhomogeneity is promoted when CO amplitude in the bulk 
 is small, and slower when the CO amplitude is large. 
We note that the generated kinks themselves show little motions~\cite{Suppl}.  
This picture is different from the case of the purely one-dimensional model 
 where a CO decrease is ascribed to solitonic motions of kinks~\cite{Hashimoto3}.
In any case, the phenomena occur in the highly nonlinear regime, 
 as seen, for example, from the threshold behavior of the induced in-gap density of states. 

\section{discussion and summary}
Finally we briefly discuss implications of our results to the experiments on PIPT, 
 although our one-dimensional model here containing only the electronic degree of freedom is too simple. 
Nevertheless, the inhomogeneity addressed above is reminiscent of the phenomena observed in molecular materials. 
As mentioned in the introduction, domain wall excitations
 and growth of the domains have been discussed in TTF-CA 
 and CO melting with domain growth in BEDT-TTF based salts. 
One discrepancy between our results here and the experimental situation is that 
 the pump light frequency needed to be tuned in the in-gap region in the simulation, 
 whereas the above-mentioned experiments use off resonance photon energy even larger than the gap.  
However, we have presented the results here with a relatively large value of $V=4$ which gives large $\Delta_\textrm{CO}$ 
 to resolve the in-gap region. 
When $V$ gets smaller, the excitations across and in the gap get considerably mixed; we show results for $V=2$ in Appendix~\ref{AppC}. 
In such cases it is possible to trigger the inhomogeneity even at large $\omega_p$. 
Another important factor is the dimensionality; higher dimensional models are known to stabilize the metallic phase at equilibrium, 
 whose dynamics are left for future problems.

In summary, we simulated the time evolution of a CO system 
 applying time-dependent HF method to a spinless fermion model. 
We found that inhomogeneity can emerge out of a homogeneous light emission 
 through cooperative effect of a collective mode excitation and the triggering by a kink structure. 


\begin{acknowledgments}
The authors would like to thank S. Iwai, A. Ono, and K. Yonemistu for valuable discussions. 
This work was supported by JSPS KAKENHI Grant
Nos. 26400377, 15H02100, 16H02393, 17H02916, and 18H05208. 
\end{acknowledgments}

\appendix

\section{Absorption spectra}\label{AppA}

\begin{figure}[b]
\begin{center}
\includegraphics[width=7.5cm]{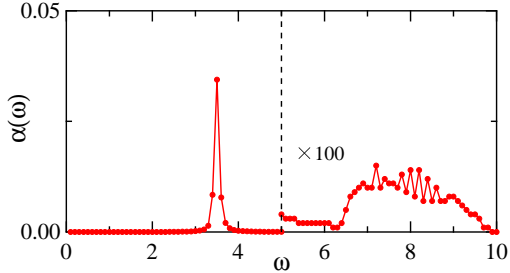}
\end{center}
\vspace{-0.8cm}
\caption{(Color online) The linear optical absorption spectrum $\alpha(\omega)$ 
using the time-dependent Hartree-Fock method, for $V=4$.}
\label{figA1}
\end{figure}

We show the optical absorption spectrum $\alpha(\omega)$, 
 calculated as the increment in the total energy upon the response to a continuous-wave gauge field 
 introduced at $\tau=0$ as $A(\tau) = A_0\ \theta(\tau) \sin(\omega\tau) \exp(-\gamma\tau) $, 
 where $\gamma$ is the damping factor. 
Figure \ref{figA1} shows the linear spectrum for the parameter set in the main text ($t=1, t'=0.1, V=4$)
 obtained for a small value of $A_0 = 0.01$ and $\gamma=0.02$. 
One can see the continuum approximately above the charge gap $\Delta_\textrm{CO} = 6.45$ 
and the resonant mode located at $\omega \sim 3.5$. 

\section{Collective mode excitations}\label{AppB}

\begin{figure}[tb]
\begin{center}
\includegraphics[width=8.8cm]{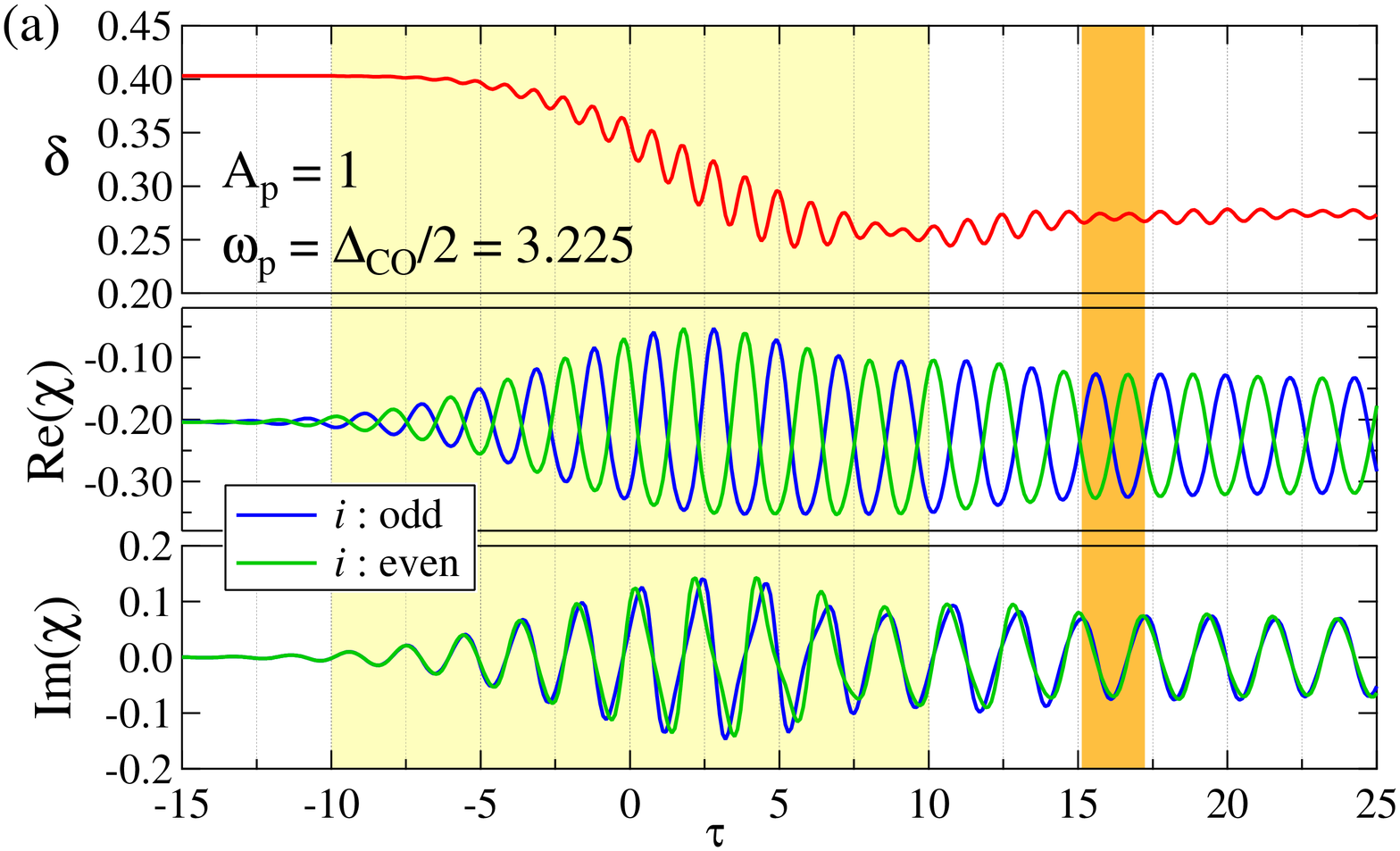}
\includegraphics[width=8.8cm]{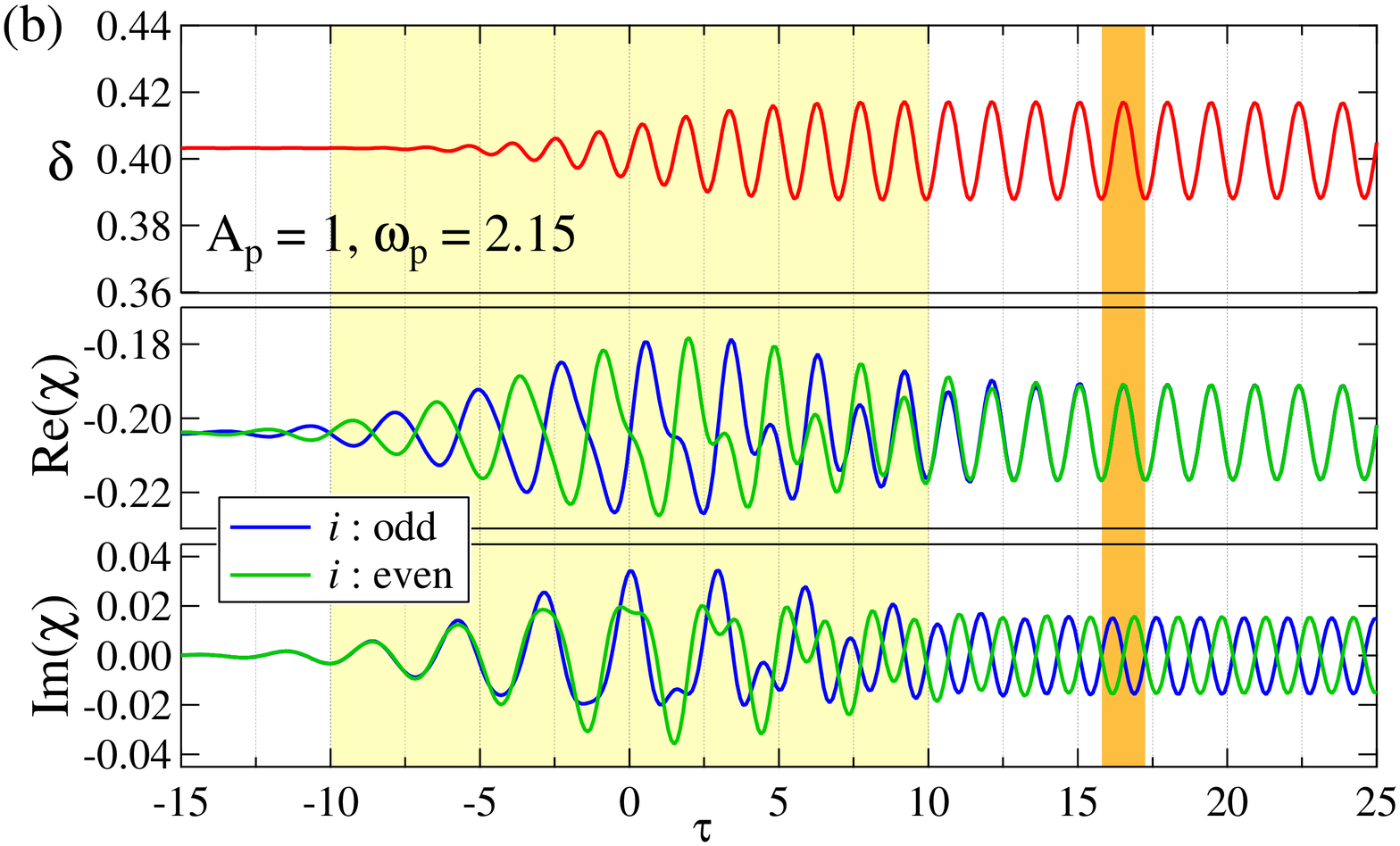}
\end{center}
\vspace{-0.5cm}
\caption{$\tau$ dependence of order parameter $\delta(\tau)$ and the real and imaginary parts of the bond orders $\chi_i(\tau)$
 for even and odd $i$ for $N=500$. 
(a) $\omega_p = \Delta_\textrm{CO}/2=3.225, A_p =1$ as in the main text [Fig.~1(b)] and 
(b) $\omega_p = 2.15, A_p = 1$. 
The light and dark shaded areas show the time domain where the pump light is applied and one period of the excited oscillation, respectively.
}
\label{figA2}
\end{figure}
We plot the time evolution of the two resonant modes observed in the simulation for $N=500$ 
 in Fig.~\ref{figA2}. 
The order parameter $\delta(\tau)$ and the real and imaginary parts of bond orders 
 $\chi_i(\tau)=\langle c_{i}^{\dagger}c_{i+1}\rangle$, for even and odd $i$, are plotted, 
 for (a) $\omega_p = \Delta_\textrm{CO}/2 =3.225, A_p=1$, and (b) $\omega_p = 2.15, A_p=1$. 
Figure \ref{figA2}(a) shows the same data as Fig.~1(b) in the main text for a longer time domain: 
 the phase mode excitation. 
Figure \ref{figA2}(b) is the amplitude mode\cite{Fukuyama}, 
 as seen in the approximately symmetric oscillation about the average $\delta_0$. 
Therefore it does not efficiently decrease CO, in contrast with the phase mode. 
The oscillation frequency is near twice of $\omega_p$, as two oscillations emerge during the light pulse; 
 the real and imaginary parts of  $\chi_i(\tau)$, the latter representing the local current, 
 show in-phase and out-of-phase alternating behavior for even and odd $i$, respectively, 
 in contrast with the phase mode. 
The phase mode excitation can be attributed to the exciton band in the linear optical spectra (see Appendix~\ref{AppA})
 discussed in the literature~\cite{Bruinsma,Gagliano,Lorenzana}, whereas the amplitude mode excitation is a two-photon process.

\section{Density of states for $N=500$}\label{AppBB}

\begin{figure}[tb]
\begin{center}
\includegraphics[width=8.8cm]{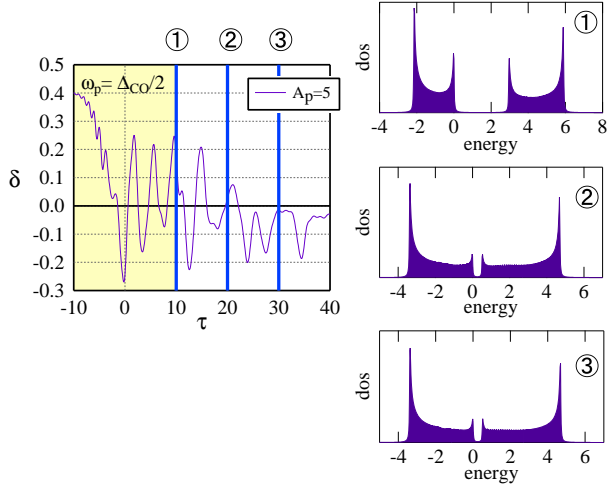}
\end{center}
\vspace{-0.5cm}
\caption{Examples of temporary density of states for the $N=500$ case, 
corresponding to the case in Fig.~\ref{fig1}(a), $\omega_p=\Delta_\textrm{CO}/2$.
The data are for $\tau=10,20,$ and 30. 
}
\label{figAppBB}
\end{figure}

In Fig.~\ref{figAppBB}, 
 we plot the temporary density of states for the $N=500$ case, 
 corresponding to the case plotted in Fig.~\ref{fig1}(a) for $\omega_p=\Delta_\textrm{CO}/2$ and $A_p=5$.
We can see that the density of states in the coherent state essentially maintains its gap structure, 
 except when the order parameter $\delta(\tau)$ coincidentally becomes 0. 
This only happens instantaneously when $\delta(\tau)$ changes sign. 
Nevertheless, in the case of $\delta(\tau)<0$ the CO just reverts it phase, 
 therefore its gap structure is common to the case of  $\delta(\tau)>0$, 
 as seen in Fig.~\ref{figAppBB}.

\section{Data for V=2}\label{AppC}
\begin{figure*}[tb]
\begin{center}
\includegraphics[width=16cm]{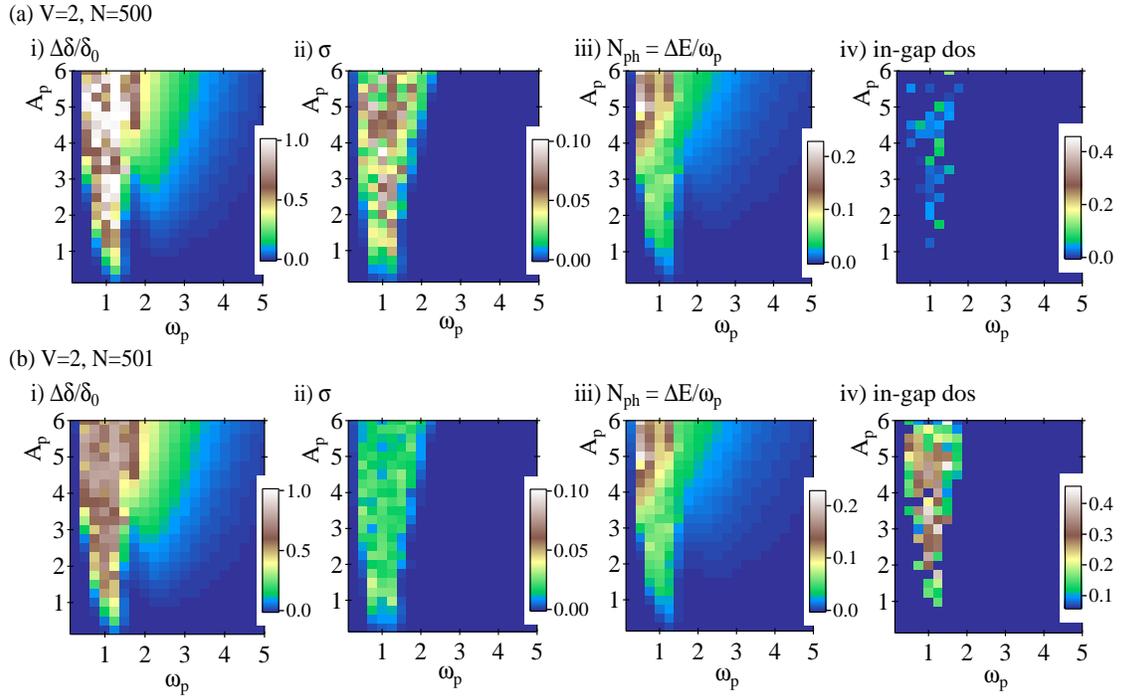}
\end{center}
\vspace{-0.7cm}
\caption{(Color online) 
Two-dimensional plots on the \{$\omega_p, A_p$\} plane of the photo-esponses for $V=2$, 
with (a) $N=500$ and (b) $N=501$ sites.  
The plotted quantities are: 
(i) decrease in the order parameter $\Delta\delta$ renormalized by $\delta_0$, 
(ii) standard deviation $\sigma$ in $\delta$, 
(iii) effective absorbed photon number $N_{ph}$, 
 and (iv) in-gap density of states (dos).
Their definitions are depicted in the main text.
}
\label{figA3}
\end{figure*}

We show two-dimensional plots of the time averaged photoresponses for the case of $V=2$, 
 as shown in the main text for $V=4$ [Figs.~\ref{fig2} and ~\ref{fig5}], 
 in Fig.~\ref{figA3} for (a) $N=500$ and (b) $N=501$. 
The plotted values are: (i)~the decrease in the order parameter $\Delta\delta$  
 renormalized by $\delta_0$, 
(ii) the standard deviation 
 $\sigma = \langle \sqrt{(\delta(\tau) - \bar{\delta})^2} \rangle_\textrm{av}$, 
 a measure for the oscillatory part, 
(iii) the effective absorbed photon number $N_{ph}$ defined as $\Delta E / \omega_p$ where $\Delta E$ 
 is the increase in the expectation value of energy, 
and (iv) the in-gap density of states (dos) measured within $\pm 0.1t$ range of the Fermi level, 
 compared to that in the $V=0$ equilibrium metallic state. 
$\Delta\delta/\delta_0$, $\sigma$, $\bar{\delta}$, and the in-gap density of states are the values averaged over $\tau=20$ to 60, 
 and $\Delta E$ is converged after the light emission in our simulation.\\ 
\indent
Similarly to the case of $V=4$, 
 the large values of $\Delta\delta/\delta_0$ are seen in the in-gap region, 
 at small $A_p$ toward $\omega_p \simeq \Delta_\textrm{CO}/2$; $\Delta_\textrm{CO}=2.05$ for $V=2$. 
The large values of in-gap dos are seen only for $N=501$ with a threshold behavior for $A_p$. 
One can see that, in contrast with the $V=4$ data in the main text, 
 the gap structure and the in-gap regime are mixed at large $A_p$ 
 and become difficult to resolve. 

%



\begin{thebibliography}{99}
\bibitem{note_randomness} 
We confirmed that introduction of very small randomness (charge densities $\simeq 10^{-5}$) 
 added to the HF self-consistent equilibrium state as the initial state does not alter the results. 
This is also the case for odd $N$.
\end{thebibliography}
\end{document}